\documentclass[prl,aps,showpacs,twocolumn,floatfix,unsortedaddress,superscriptaddress]{revtex4-1}
\usepackage{graphics, bm, psfrag, amsmath, amssymb, epsfig, grffile, float, bbold}
\usepackage{subfigure}
\usepackage{dsfont}
\usepackage{color}
\usepackage{hyperref}
\usepackage{natbib} 

\begin{document}

\title{Odd-frequency Pairing in Conventional Josephson Junctions}

\author{Alexander V. Balatsky}
\affiliation{Institute for Materials Science (IMS), Los Alamos National Laboratory, Los Alamos New Mexico 87545, USA}
\affiliation{Nordic Institute for Theoretical Physics (NORDITA), Stockholm, Sweden}
\affiliation{Center for Quantum Materials (CQM), KTH and Nordita, Stockholm, Sweden}

\author{Sergey S. Pershoguba}
\affiliation{Nordic Institute for Theoretical Physics (NORDITA), Stockholm, Sweden}
\affiliation{Center for Quantum Materials (CQM), KTH and Nordita, Stockholm, Sweden}
\affiliation{Department of Physics, Yale University, New Haven, Connecticut 06520, USA}

\author{Christopher Triola}
\affiliation{Nordic Institute for Theoretical Physics (NORDITA), Stockholm, Sweden}
\affiliation{Center for Quantum Materials (CQM), KTH and Nordita, Stockholm, Sweden}
\affiliation{Department of Physics and Astronomy, Uppsala University, Box 516, S-751 20 Uppsala, Sweden}

\begin{abstract}
Using a simple theoretical model, we demonstrate the emergence of odd-frequency pair amplitudes in conventional Josephson junctions both in the absence of a voltage (DC effect) and in the presence of a finite voltage (AC effect). In both cases, we find that odd-frequency interlead pairing emerges whenever a Josephson current is expected to flow.
Additionally, we show that the interlead spin-susceptibility is directly influenced by the presence of the odd-frequency pair amplitudes. Specifically, we find that the spin-susceptibility is suppressed when the odd-frequency component is the largest. By establishing a novel link between the physics of Josephson junctions and odd-frequency pairing, this work demonstrates the importance of odd-frequency pairing for understanding conventional superconducting systems.
\end{abstract}

\maketitle
\paragraph*{Introduction---}

Odd-frequency (odd-$\omega$) pairing, originally posited by Berezinskii \cite{Berezinskii1974} in the context of superfluid $^3$He and later extended to superconductivity\cite{kirkpatrick1991,belitz1992,BalatskyPRB1992}, refers to the possibility that the fermionic pairing function describing a condensate, $F(\tau_1,\tau_2) = -\langle T_\tau \psi(\tau_1) \psi(\tau_2)  \rangle$, is odd under the interchange of $\tau_1$ and $\tau_2$, or, equivalently, odd in Matsubara frequency. Unlike conventional superconductors, which only allow for pair symmetries which are either spin-singlet and even-parity ($s$-wave, $d$-wave, etc.) or spin-triplet and odd-parity ($p$-wave, $f$-wave, etc.), odd-$\omega$ pairing allows for a wider variety of pair symmetries like: spin-singlet $p$-wave or spin-triplet $s$-wave states. In addition to opening the door for exotic pair symmetries, odd-$\omega$ pairing represents a class of hidden order due to the vanishing of equal time correlations \cite{linder2017odd}.

Following the initial proposal for odd-$\omega$ superconductivity, several studies have been conducted dedicated to the thermodynamic stability of intrinsically odd-$\omega$ phases \cite{heid1995thermodynamic,solenov2009thermodynamical,kusunose2011puzzle,FominovPRB2015}. A growing list of systems expected to host this unusual pairing state have been  identified, including: ferromagnet-superconductor heterostructures \cite{BergeretPRL2001,bergeret2005odd,yokoyama2007manifestation,houzet2008ferromagnetic,EschrigNat2008,LinderPRB2008,crepin2015odd}, topological insulator-superconductor systems \cite{YokoyamaPRB2012,Black-SchafferPRB2012,Black-SchafferPRB2013,TriolaPRB2014,cayao2017prb}, normal metal-superconductor junctions \cite{tanaka2007theory,TanakaPRB2007,LinderPRL2009,LinderPRB2010_2,TanakaJPSJ2012}, two-dimensional electron systems coupled to bulk superconductors \cite{parhizgar_2014_prb,triola2016prl}, multiband superconductors with a finite interband hybridization \cite{black2013odd,komendova2015experimentally,komendova2017odd,triola2018prb}, and conventional superconductors subjected to time-dependent drives \cite{triolaprb2016,triola2017pair}. In addition to theoretical studies, there are experimental indications of the realization of odd-$\omega$ pairing at the interface of Nb thin films and epitaxial Ho \cite{di2015signature,di2015intrinsic}. Furthermore, the concept of odd-$\omega$ order parameters can be generalized to charge and spin-density waves \cite{pivovarov2001odd,kedem2015odd} and Majorana fermion pairs \cite{huang2015odd}.
\begin{figure}
 \includegraphics[width=0.8\linewidth]{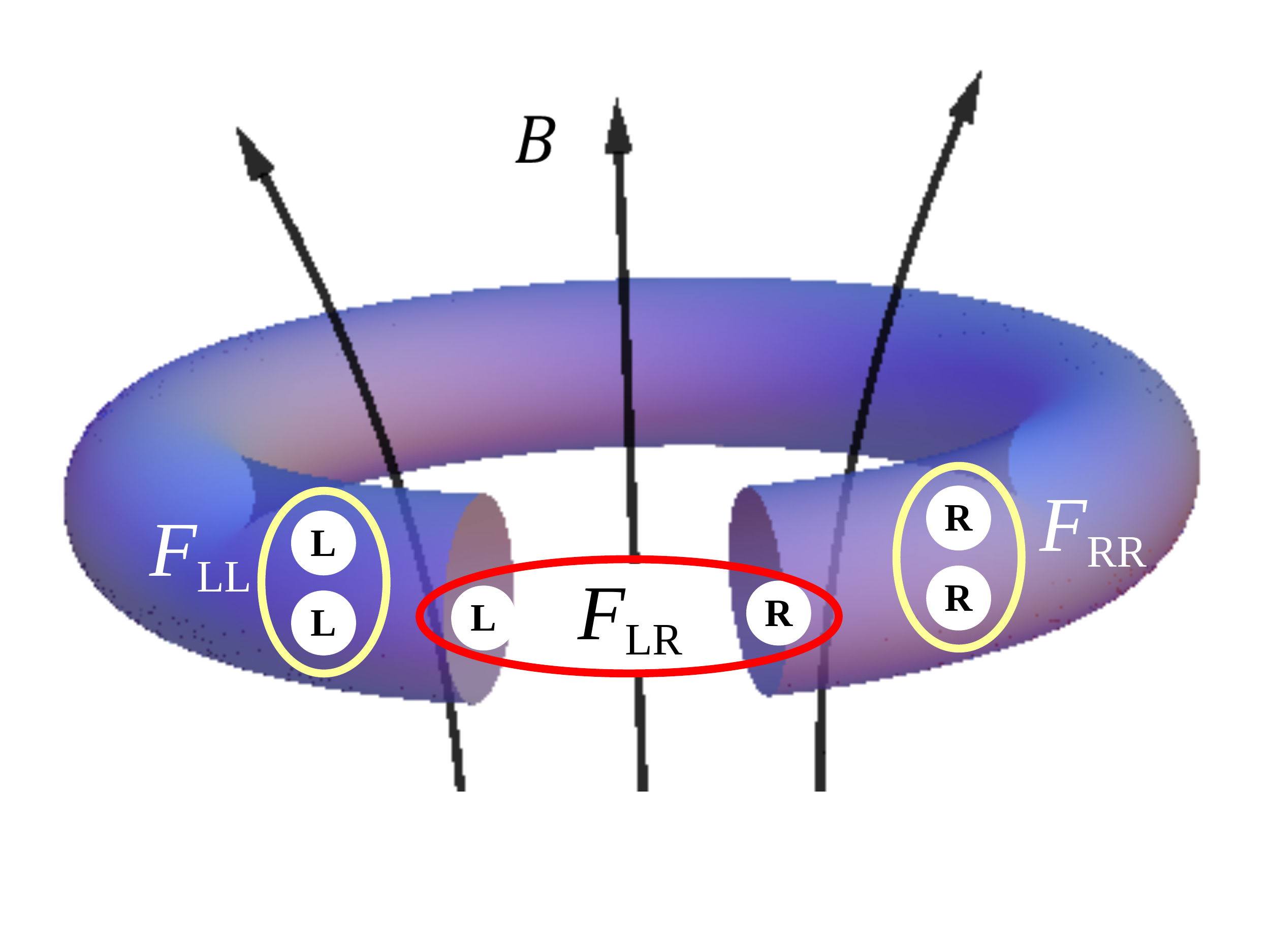}
 \caption{Josephson junction in a ring geometry. The phase difference $\phi_{LR} = \phi_L-\phi_R$ between the two superconducting leads is controlled by an external magnetic field $B$. If the phase is non-zero $\phi_{LR}\neq 0$, dissipationless Josephson current flows across the junction in the form of intralead Cooper pairs, $F_{LL/RR} = \langle\psi_{L/R} \psi_{L/R}\rangle$. At the same time, odd-frequency interlead pairing $F_{LR} = \langle\psi_L \psi_R\rangle$ is induced.}
\label{fig1}
\end{figure}

Keeping in mind the ubiquity of odd-frequency states, we revisit a textbook example of a classical superconducting Josepshon junction (JJ) illustrated in Fig.~\ref{fig1}. It is well-known that a dissipationless Josepshon current
\begin{equation}
	I=I_c\sin\phi_\text{LR} \label{jjDC}
\end{equation}
flows between the leads labelled here by ``L'' and ``R'' in response to the phase difference $\phi_{\rm LR} = \phi_{\rm L}-\phi_{\rm R}$ controlled externally. Furthermore, if there is a finite voltage drop $V$ across the junction, the phase difference $\phi_{\rm LR}$ develops a time dependence

\begin{equation}
	V=\frac{\hbar}{2e}\frac{\partial \phi_\text{LR}(t)}{\partial t}. \label{jjAC}
\end{equation}
Taken together, Eqs.~(\ref{jjDC}) and (\ref{jjAC}) describe the essential phenomenology of both the DC and AC Josephson effects.

In this paper we will demonstrate a previously overlooked feature of classical Josephson junctions: that {\it emergent odd-$\omega$ interlead superconducting correlations always accompany the flow of a Josephson current}, as depicted in Fig.~\ref{fig1}. An intuitive way to see that this odd-$\omega$ cross-junction pairing might arise is to consider that the anomalous Green's function for the system, $F$, must satisfy the Berezinskii condition: $\mathcal S \mathcal T \mathcal O \mathcal P F=-F$ \cite{triolaprb2016,linder2017odd}, where $\mathcal S$, $\mathcal T$, $\mathcal O$, and $\mathcal P$ are operators corresponding to the exchange of the spin, time, orbital, and spatial indices, respectively. In the case of a conventional JJ, pairing is on-site and spin-singlet, which corresponds to $\mathcal P=+1$ and $\mathcal S=-1$, and, the orbital index $\mathcal O$ labels the two distinct leads L and R. Thus, the Berezinskii constraint requires $\mathcal O \mathcal T=+1$. Therefore, two kinds of superconducting correlations are allowed in conventional JJs: correlations which are orbital-even $\mathcal O = +1$ and frequency-even $\mathcal T = +1$; and correlations which are orbital-odd $\mathcal O = -1$ and frequency-odd $\mathcal T=-1$. In both cases, the product $\mathcal O \mathcal T=+1$ remains fixed as demanded by the Berezinskii condition. Clearly, the diagonal components, $F_{\text{LL}},F_{\text{RR}}$, must lie in the class $\mathcal{O},\mathcal{T}=+1$, while the off-diagonal components, $F_{\text{LR}},F_{\text{RL}}$, could, in general, possess terms lying in either class. However, as we will show, these interlead corrections are strictly odd in frequency at the interface. Furthermore, the conventional JJ coupling, $t_0$, results in a quadratic correction to the diagonal components $F_{\rm LL},F_{\rm RR} \sim t_0^2$, whereas the leading-order terms in the interlead pairing are linear in $t_0$, $F_{\rm LR} \sim t_0$. Hence, the dominant corrections to the anomalous Green's functions in a conventional Josephson junction are precisely those that are odd in frequency.

\paragraph{Odd-frequency in the DC Josephson regime ---} \label{sec:dc}
To gain some insight into the properties of the interlead superconducting pair amplitudes in a JJ, we will start from a simple model describing a system similar to the structure illustrated in Fig.~\ref{fig1}. We model the two-superconducting leads ``L'' and ``R'' as two independent superconductors coupled by a point contact. The corresponding Hamiltonian takes the form $H=H_{0}+H_T$ where
\begin{equation}
\begin{aligned}
& H_0=\sum_{\textbf{k},\sigma,\alpha}\xi_{\textbf{k},\alpha} \psi^\dagger_{\textbf{k},\sigma,\alpha}\psi_{\textbf{k},\sigma,\alpha} \\
 & + \Delta_0 \sum_{\textbf{k},\alpha}  e^{i\phi_\alpha} \psi^\dagger_{-\textbf{k},\uparrow,\alpha}\psi^\dagger_{\textbf{k},\downarrow,\alpha} + e^{-i\phi_\alpha} \psi_{\textbf{k},\downarrow,\alpha}\psi_{-\textbf{k},\uparrow,\alpha}, \\
 &H_{T}=\frac{t_0}{\mathcal{V}}\sum_{\textbf{k},\textbf{k}',\sigma} \psi^\dagger_{\textbf{k},\sigma,\text{L}}\psi_{\textbf{k}',\sigma,\text{R}} + \text{h.c.}
\end{aligned}
\label{eq:Hsc}
\end{equation}
Here $\psi^\dagger_{\textbf{k},\sigma,\alpha}$ ($\psi_{\textbf{k},\sigma,\alpha}$) creates (annihilates) a quasiparticle state with momentum $\textbf{k}$ and spin $\sigma$ in the superconducting lead indexed by $\alpha\in\{\text{L},\text{R}\}$, $\xi_{\textbf{k},\alpha}$ is the normal state quasiparticle dispersion of the superconducting leads measured from the chemical potential $\mu_\alpha$, $\Delta_0$ is the magnitude of the order parameter of the two superconductors, assumed to be equal for both L and R, $\phi_\alpha$ is the complex phase of the superconducting order parameter on each side of the junction, and $t_0$ parameterizes the local tunneling across the junction at position $\textbf{r}=0$, and $\mathcal{V}$ is an effective volume of each of the leads which are assumed equal.

In the absence of tunneling, $t_0=0$, it is straightforward to write down the Matsubara Green's functions for the two superconductors
\begin{equation}
\begin{aligned}
	G^{(0)}_{\alpha\beta}(\textbf{k}_1,\textbf{k}_2;i\omega_n)&=-\delta_{\bm k_1,\bm k_2}\delta_{\alpha,\beta}\frac{i\omega_n + \xi_{\textbf{k}_1}}{\omega_n^2 + \xi_{\textbf{k}_1}^2 + \Delta_0^2}, \\
	F^{(0)}_{\alpha\beta}(\textbf{k}_1,\textbf{k}_2;i\omega_n)&=-\delta_{\bm k_1,\bm k_2}\delta_{\alpha,\beta}\frac{\Delta_0e^{i\phi_{\alpha}}}{\omega_n^2 + \xi_{\textbf{k}_1}^2 + \Delta_0^2}, \\
\end{aligned}
\label{eq:g_0}
\end{equation}
where $\alpha,\beta\in\{\text{L},\text{R}\}$ label the superconducting leads, and we explicitly keep track of the two momenta, $\textbf{k}_1$ and $\textbf{k}_2$, since the presence of the junction at $\textbf{r}=0$ breaks spatial translation-invariance.

In the presence of a weak interlead tunneling $t_0$ we can evaluate corrections to the Green's functions perturbatively in $t_0$. Since the tunneling is local in real space it scatters states in lead ``L" to states in lead ``R" without conserving momentum. Thus, the first order corrections in the perturbative expansion of $F$ are given by
\begin{align}
	F^{(1)}_{\text{R}\text{L}}(\textbf{k}_1, & \textbf{k}_2;i\omega_n) = \label{eq:f_lr0}\\
 & \frac{t_0}{\mathcal{V}}\sum_{\textbf{k},\textbf{k}''}\left[G^{(0)}_{\text{R}\text{R}}(\textbf{k}_1,\textbf{k}';i\omega_n)F^{(0)}_{\text{L}\text{L}}(\textbf{k}'',\textbf{k}_2;i\omega_n) \right., \nonumber \\
 & +\left.F^{(0)}_{\text{R}\text{R}}(\textbf{k}_1,\textbf{k}';i\omega_n)G^{(0)}_{\text{L}\text{L}}(\textbf{k}'',\textbf{k}_2;i\omega_n)^* \right]. \nonumber	
\end{align}
Inserting the expressions for the Green's functions from Eqs (\ref{eq:g_0}) we see that the linear corrections to the anomalous Green's functions are given by
\begin{align}
	F^{(1)}_{\text{R}\text{L}}(\textbf{k}_1, & \textbf{k}_2;i\omega_n) = \label{eq:f_lr}\\
& \frac{t_0}{\mathcal{V}}\dfrac{\Delta_0\left[e^{i\phi_\text{R}}\left(i\omega_n+ \xi_{\textbf{k}_1}\right) - e^{i\phi_\text{L}}\left(i\omega_n- \xi_{\textbf{k}_2}\right)\right]}{\left[ \omega_n^2 +\xi_{\textbf{k}_1}^2+\Delta_0^2 \right]\left[ \omega_n^2 +\xi_{\textbf{k}_2}^2+\Delta_0^2 \right]}. \nonumber
\end{align}
Notice from Eq. (\ref{eq:f_lr}) that this component of the anomalous Green's function now contains both even-$\omega$ and odd-$\omega$ terms. In order to expose the odd-$\omega$ term, we evaluate the on-site anomalous Green's function at the junction, i.e. $\bm r = 0$, by summing over the independent momenta $\bm k_1$ and $\bm k_2$
\begin{align}
&	F^{(1)}_{\rm RL}(\bm r=0;i\omega_n) = \dfrac{2\omega_n\pi^2 \rho^2 t_0 \Delta_0 e^{i\tfrac{\phi_\text{R}+\phi_\text{L}}{2}} \sin\frac{\phi_\text{LR}}{2}}{ \omega_n^2 +\Delta_0^2}. \label{eq:f_odd}
\end{align}
where $\rho$ is the quasiparticle density of states at the chemical potential. From Eq (\ref{eq:f_odd}) we can see that the odd-$\omega$ pairing is the only interlead channel appearing at $\bm r =0$ to this order in the tunneling. Furthermore, we note that the odd-$\omega$ component is non-zero as long as $\phi_{\text{LR}} \neq 2\pi n$.
We can also evaluate the time dependence of the RL coherence set by the tunneling. After Wick-rotating Eq (\ref{eq:f_odd}) to real-time ($\tau\rightarrow it$) and taking the zero-temperature limit, we find that 
\begin{equation}
F^{(1)}_{\rm RL}(\bm r=0; t) = -i\pi^2\rho^2 t_0 \Delta_0 e^{i\tfrac{\phi_\text{R}+\phi_\text{L}}{2}} \sin\tfrac{\phi_\text{LR}}{2}e^{-i\Delta_0t}.
\end{equation} 
These time-dependent oscillations of the real and imaginary parts of the RL amplitude are present as long as one maintains the phase difference across the junction, in a fashion reminiscent of the Rabi oscillations \cite{linder2017odd}.

Within this model, the Josephson current is given by
\begin{align}
	I_{\rm Josephson} \propto t_0^2 \rho^2 \pi^2\Delta \sin \phi_{\text{LR}},
	\label{jcur}
\end{align}
which can also be obtained perturbatively in $t_0$ (see Supplemental Materials\cite{sm} for details). Comparing this expression for the Josephson current to the expression for the magnitude of the odd-$\omega$ pairing, Eq.~(\ref{eq:f_odd}), we notice two key similarities: both are periodic functions of $\phi_{\text{LR}}$; and both are non-zero for generic values of $\phi_{\text{LR}}$ but vanish at particular integer multiples of $\pi$. However, in contrast to Eq.~(\ref{eq:f_odd}) the Josephson current vanishes for $\phi_{\text{LR}} = \pi n$, while the odd-$\omega$ pair amplitude vanishes for $\phi_{\text{LR}} = 2\pi n$. Hence, for $\phi_{\text{LR}} = \pi (2n+1)$ we see that the Josephson current will vanish but the odd-$\omega$ pair amplitude will reach its maximum value. Nevertheless, whenever a Josephson current flows, odd-$\omega$ pairing will be present in the system. 
\begin{figure}
 \includegraphics[width=0.9\linewidth]{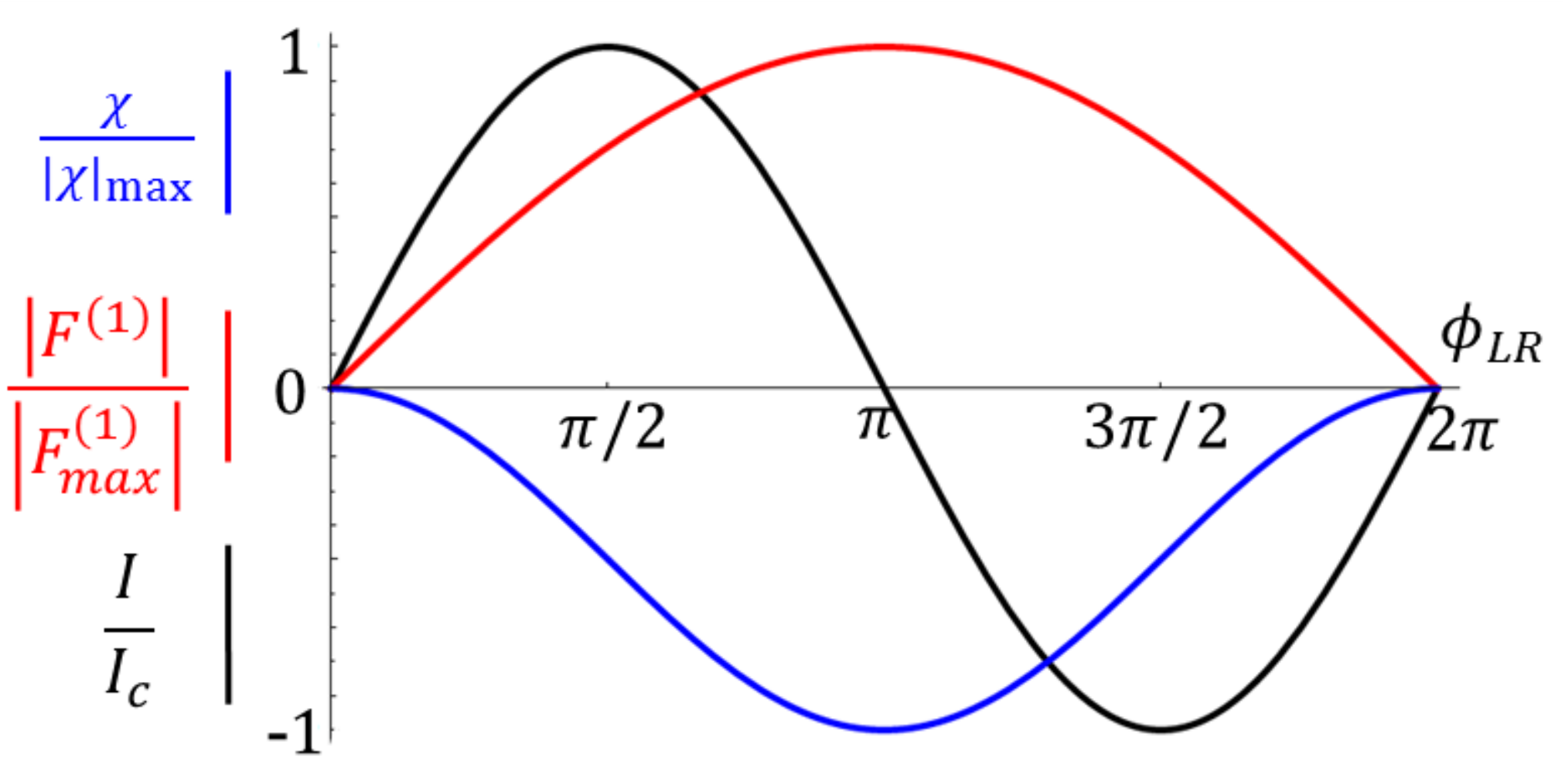}
 \caption{Dependence on the Josephson phase, $\phi_{\text{LR}}$, of: the odd-frequency amplitude, Eq~(\ref{eq:f_odd}), (red); Josephson current, Eq~(\ref{jcur}), (black); and interlead spin-susceptibility, Eq~(\ref{ss1}), (blue). For the spin-susceptibility, only the contribution due to the anomalous terms, $F^{(1)}$, which are odd in $\omega$, was used.}
\label{fig3}
\end{figure}

\paragraph*{Interlead spin-susceptibility---}
Above we demonstrated that the Josephson current itself can be used as a signature for the presence of odd-$\omega$ pairing. However, this signature is somewhat indirect since the two are only correlated and not necessarily dependent on one another. Now we will investigate another observable that is directly dependent on the presence of the odd-$\omega$ pair amplitudes, the interlead spin susceptibility $\chi$.

In a conventional s-wave superconductor the Cooper pairs exist in a spin singlet state and are therefore overall spinless. This leads to an exponentially suppressed on-site spin-susceptibility at low temperatures~\cite{Yosida1958}. However, the spin-susceptibility does not vanish at a finite distance $\bm r$ because the Cooper pairs have a finite size. Formally, the interlead spin-susceptibility $\chi_{\alpha\beta}$ determines the interaction energy $U = \chi \bm S_L\cdot \bm S_R$ between the spins localized on the different leads $\bm S_\alpha \psi^\dagger_{\alpha \sigma} \bm \sigma_{\sigma\sigma'} \psi_{\alpha\sigma'}$, where $\alpha = \text{L},\text{R}$. The contribution to the spin-susceptibility due to the anomalous Green's function is given by,
\begin{align}
	\chi =  -2T \sum_{i\omega_n}\left[ F^{(1)}_{\text{LR}}(\bm r = 0;i\omega_n)F^{(1)}_{\text{LR}}(\bm r = 0;i\omega_n)^*\right]. \label{ss}
\end{align}
In contrast to the expression for the free energy\cite{sm}, the spin-susceptibility, Eq.~(\ref{ss}), is constructed from the first-order interlead Green's functions. In particular, Eq.~(\ref{ss}) consists of purely odd-$\omega$ anomalous propagators $F^{(1)}$. Evaluation of Eq~(\ref{ss}) at $T=0$ gives
\begin{equation}
\begin{aligned}
	\chi &= -2\pi^4\rho^4 t_0^2\Delta \sin^2\frac{\phi_{\text{LR}}}{2},
\end{aligned}
\label{ss1}
\end{equation}
Notice that the spin-susceptibility  is {\em finite, nonexponential in T, and the largest in magnitude at $\phi_{\rm LR} = \pi$ }, i.e. where the odd-$\omega$ pair amplitude is the largest. In this way, the spin-susceptibility could be used as an observable signature of the odd-frequency terms.

In Fig.~\ref{fig3} we summarize the results, by plotting the odd-frequency amplitude from Eq (\ref{eq:f_odd}), the Josephson current, Eq. (\ref{jcur}), and the spin-susceptibility, Eq.~(\ref{ss1}), with respect to the JJ phase difference $\phi_{\text{LR}}$. All quantities have distinct $\phi_{\text{LR}}$ dependencies but exist within one device and can be controlled by an external magnetic flux.

\paragraph*{Odd-frequency in the AC Josephson regime---} \label{sec:ac}
We turn our attention to the case in which the chemical potentials in the two superconducting leads are not equal, instead we have $\mu_\text{L}-\mu_\text{R}=\Delta\mu$. This corresponds to applying a finite voltage across the Josephson junction, $V=-\Delta\mu/e$. In this case, from Eq. (\ref{jjAC}), we observe that the complex phase difference, $\phi_\text{LR}(t)$, is given by:
\begin{equation}
\phi_{\text{LR}}(t)=\Delta\phi_0-\Omega_{\text{J}} t,
\label{eq:phi_t}
\end{equation}
where $\Omega_{\text{J}}=2\Delta\mu/\hbar$ and $\Delta\phi_0$ is the initial phase difference across the junction.

We note that the system is still described by the Hamiltonian in Eq. (\ref{eq:Hsc}) except the phases $\phi_\alpha$ now have time dependence:
\begin{equation}
\begin{aligned}
\phi_{\text{L,R}}(t)&=\Phi_0 \pm \frac{\phi_{\text{LR}}(t)}{2},
\end{aligned}
\label{eq:phi_alpha}
\end{equation}
where $\Phi_0$ is the average phase of the two gaps, $\Phi_0=(\phi_\text{L}+\phi_\text{R})/2$.

For convenience, we will perform a gauge transformation on the fermionic fields:
\begin{equation}
\begin{aligned}
\psi_{\textbf{k},\sigma,\alpha}&\rightarrow \tilde{\psi}_{\textbf{k},\sigma,\alpha}= e^{i\frac{\phi_{\alpha}(t)}{2}} \psi_{\textbf{k},\sigma,\alpha}(t), \\
\mu_\alpha&\rightarrow \mu_\alpha+\frac{\hbar}{2}\frac{\partial \phi_{\alpha}(t)}{\partial t}.
\end{aligned}
\label{eq:gaugetransformation}
\end{equation}
After this transformation, the Hamiltonian becomes $H_t=H_{0}'+H'_{T}(t)$ where $H_0'$ is identical to $H_0$ from Eq (\ref{eq:Hsc}) with the left and right chemical potential replaced by an average chemical potential $\mu_\text{L}'=\mu_\text{R}'=\mu\equiv(\mu_\text{L}+\mu_\text{R})/2$. While the tunneling Hamiltonian develops a time-dependence $H'_{T}(t)$ given by
\begin{equation}
\begin{aligned}
H'_T(t)&=\frac{t_0}{\mathcal{V}}\sum_{\textbf{k},\textbf{k}',\sigma}  e^{-i\frac{\phi_{\text{LR}}(t)}{2}} \tilde{\psi}^\dagger_{\textbf{k},\sigma,\text{L}}(t)\tilde{\psi}_{\textbf{k}',\sigma,\text{R}}(t) +\text{h.c.}.
\end{aligned}
\label{eq:H_alpha_t}
\end{equation}

In the absence of tunneling ($t_0=0$) we can see that the system is described by the Hamiltonian in Eq.~(\ref{eq:Hsc}) with $t_0=0$ and $\phi_\text{L}=\phi_\text{R}=0$. In this limit the frequency space representation of the retarded Green's functions may be obtained by analytically continuing the expressions in Eq.~(\ref{eq:g_0}) using the usual prescription $i\omega_n\rightarrow \omega +i0^{+}$.

Turning our attention to the case of finite tunneling ($t_0\neq 0$) we can evaluate the leading order corrections to the anomalous Green's function given by
\begin{widetext}
\begin{equation}
\begin{aligned}
F^{(1)}_{\text{R}\text{L}}(\textbf{k}_1,  \textbf{k}_2;\omega,t_{\text{av}}) &= \sum_{\textbf{k},\textbf{k}''}\int \frac{d\Omega}{2\pi}e^{-it_{\text{av}}\Omega}\left[G^{(0)}_{\text{R}\text{R}}(\textbf{k}_1,\textbf{k}';\omega+\tfrac{\Omega}{2})T(\Omega)F^{(0)}_{\text{L}\text{L}}(\textbf{k}'',\textbf{k}_2;\omega-\tfrac{\Omega}{2}) \right. \\
&+\left.F^{(0)}_{\text{R}\text{R}}(\textbf{k}_1,\textbf{k}';\omega+\tfrac{\Omega}{2})T^*(-\Omega)G^{(0)}_{\text{L}\text{L}}(\textbf{k}'',\textbf{k}_2;-\omega+\tfrac{\Omega}{2})^* \right],
\end{aligned}
\label{eq:flrt}
\end{equation}
\end{widetext}
where $\omega$ is the frequency associated with the relative time, $t_\text{rel}=t_1-t_2$, we define the average time $t_{\text{av}}=(t_1+t_2)/2$, and the frequency-dependent tunneling is given by
\begin{equation}
T(\omega)=2\pi \frac{t_0}{\mathcal{V}}e^{-i\frac{\Delta\phi_{0}}{2}}\delta(\omega+\tfrac{\Omega_\text{J}}{2}),
\end{equation}
which can be obtained by Fourier transforming the expression in Eq (\ref{eq:H_alpha_t}).

We evaluate Eq (\ref{eq:flrt}) at the interface $\bm r=0$ by integrating over the two independent momenta, $\textbf{k}_1$, $\textbf{k}_2$, to find
\begin{equation}
\begin{aligned}
F^{(1)}_{\text{RL}}(\bm r=0;\omega,t_{\text{av}})&=\frac{i2t_0\Delta_0\rho^2\pi^2\sin\left(\frac{\Omega_\text{J}t_{\text{av}}-\Delta\phi_{0}}{2}\right)\omega_{-}}{\sqrt{\Delta_0^2-\omega_{-}^2}\sqrt{\Delta_0^2-\omega_{+}^2}},
\end{aligned}
\label{eq:fr_lr_fermi}
\end{equation}
where we define $\omega_{\pm}=\omega\pm\frac{\Omega_\text{J}}{4}$ and where $\rho$ is the density of states at the Fermi level.

Notice, in the static limit ($\Omega_\text{J}\rightarrow 0$) only the odd-$\omega$ term remains and that it is exactly the analytic continuation of the term in Eq. (\ref{eq:f_odd}). In the case of an AC junction we have an additional even-$\omega$ term which scales as the frequency of the Josephson current. This result bears some similarity to the result of a recent work \cite{triola2017pair} in which we showed that the presence of a time-dependent drive can convert both even-$\omega$ pairing to odd-$\omega$ pairing and vice versa. In this case, the role of the time-dependent drive is played by the time-dependent phase.

Comparing Eq. (\ref{eq:fr_lr_fermi}) to the expression for the Josephson current in this system, we see that they share similar dependence on average time, $t_{\text{av}}$. While the Josephson current goes as $\sin\left(\Omega_\text{J}t_{\text{av}}-\Delta\phi_{0}\right)$, the odd-$\omega$ pair amplitude goes as $\sin\left(\tfrac{\Omega_\text{J}t_{\text{av}}-\Delta\phi_{0}}{2}\right)$. Therefore, we conclude that, whenever there is a Josephson current flowing across the junction, odd-$\omega$ interlead pairing will also exist in the junction.

\paragraph*{Concluding remarks---} \label{sec:con}

In this work we considered a simple Josephson junction comprised of two superconducting regions coupled via a weak link. In this well-known system, we investigated the pair symmetry in the interlead channel and found significant odd-frequency pair amplitudes. We considered the cases of DC and AC Josephson effects and found that the presence of a finite Josephson current was always accompanied by odd-frequency interlead pairing. This finding underscores two key aspects of the odd frequency pairing: i) it is an example of hidden order - to the best of our knowledge, the presence of the odd-$\omega$ pairs was not reported before; and ii) our work adds conventional Josephson junctions to the list of systems hosting odd-frequency pairing, thus demonstrating the ubiquity of the latter in fairly mundane superconducting systems. We also computed an interlead spin-susceptibility $\chi$ perturbatively in the tunneling and found it to be finite at low temperatures and directly connected to the odd-frequency pairing amplitude. Therefore, we propose Josephson junctions as a convenient experimental platform where the magnitude of the odd-frequency component can be tuned by changing the Josephson phase $\phi_{\text{LR}}$. Our finding significantly expands the conversation about superconducting coherence effects in Josephson Junctions, a subject of continued interest since the pioneering work of Josephson.

Acknowledgements: We wish to thank Annica Black-Schaffer, Jorge Cayao, Matthias Geilhufe, Yaron Kedem, Jacob Linder, and Fariborz Parhizgar for useful discussions. This work was supported by  KAW and the European Research Council (ERC) DM-321031.

\bibliographystyle{apsrev}
\bibliography{Odd_Frequency}

\begin{thebibliography}{41}
\expandafter\ifx\csname natexlab\endcsname\relax\def\natexlab#1{#1}\fi
\expandafter\ifx\csname bibnamefont\endcsname\relax
  \def\bibnamefont#1{#1}\fi
\expandafter\ifx\csname bibfnamefont\endcsname\relax
  \def\bibfnamefont#1{#1}\fi
\expandafter\ifx\csname citenamefont\endcsname\relax
  \def\citenamefont#1{#1}\fi
\expandafter\ifx\csname url\endcsname\relax
  \def\url#1{\texttt{#1}}\fi
\expandafter\ifx\csname urlprefix\endcsname\relax\def\urlprefix{URL }\fi
\providecommand{\bibinfo}[2]{#2}
\providecommand{\eprint}[2][]{\url{#2}}

\bibitem[{\citenamefont{Berezinskii}(1974)}]{Berezinskii1974}
\bibinfo{author}{\bibfnamefont{V.~L.} \bibnamefont{Berezinskii}},
  \bibinfo{journal}{Pis' ma Zh. Eksp. Teor. Fiz.}
  \textbf{\bibinfo{volume}{20}}, \bibinfo{pages}{628} (\bibinfo{year}{1974}).

\bibitem[{\citenamefont{Kirkpatrick and Belitz}(1991)}]{kirkpatrick1991}
\bibinfo{author}{\bibfnamefont{T.~R.} \bibnamefont{Kirkpatrick}}
  \bibnamefont{and} \bibinfo{author}{\bibfnamefont{D.}~\bibnamefont{Belitz}},
  \bibinfo{journal}{Phys. Rev. Lett.} \textbf{\bibinfo{volume}{66}},
  \bibinfo{pages}{1533} (\bibinfo{year}{1991}),
  \urlprefix\url{https://link.aps.org/doi/10.1103/PhysRevLett.66.1533}.

\bibitem[{\citenamefont{Belitz and Kirkpatrick}(1992)}]{belitz1992}
\bibinfo{author}{\bibfnamefont{D.}~\bibnamefont{Belitz}} \bibnamefont{and}
  \bibinfo{author}{\bibfnamefont{T.~R.} \bibnamefont{Kirkpatrick}},
  \bibinfo{journal}{Phys. Rev. B} \textbf{\bibinfo{volume}{46}},
  \bibinfo{pages}{8393} (\bibinfo{year}{1992}),
  \urlprefix\url{https://link.aps.org/doi/10.1103/PhysRevB.46.8393}.

\bibitem[{\citenamefont{Balatsky and Abrahams}(1992)}]{BalatskyPRB1992}
\bibinfo{author}{\bibfnamefont{A.}~\bibnamefont{Balatsky}} \bibnamefont{and}
  \bibinfo{author}{\bibfnamefont{E.}~\bibnamefont{Abrahams}},
  \bibinfo{journal}{Phys. Rev. B} \textbf{\bibinfo{volume}{45}},
  \bibinfo{pages}{13125} (\bibinfo{year}{1992}),
  \urlprefix\url{https://link.aps.org/doi/10.1103/PhysRevB.45.13125}.

\bibitem[{\citenamefont{Linder and Balatsky}(2017)}]{linder2017odd}
\bibinfo{author}{\bibfnamefont{J.}~\bibnamefont{Linder}} \bibnamefont{and}
  \bibinfo{author}{\bibfnamefont{A.~V.} \bibnamefont{Balatsky}},
  \bibinfo{journal}{arXiv preprint arXiv:1709.03986}  (\bibinfo{year}{2017}).

\bibitem[{\citenamefont{Heid}(1995)}]{heid1995thermodynamic}
\bibinfo{author}{\bibfnamefont{R.}~\bibnamefont{Heid}},
  \bibinfo{journal}{Zeitschrift f{\"u}r Physik B Condensed Matter}
  \textbf{\bibinfo{volume}{99}}, \bibinfo{pages}{15} (\bibinfo{year}{1995}),
  ISSN \bibinfo{issn}{1431-584X},
  \urlprefix\url{https://doi.org/10.1007/s002570050003}.

\bibitem[{\citenamefont{Solenov et~al.}(2009)\citenamefont{Solenov, Martin, and
  Mozyrsky}}]{solenov2009thermodynamical}
\bibinfo{author}{\bibfnamefont{D.}~\bibnamefont{Solenov}},
  \bibinfo{author}{\bibfnamefont{I.}~\bibnamefont{Martin}}, \bibnamefont{and}
  \bibinfo{author}{\bibfnamefont{D.}~\bibnamefont{Mozyrsky}},
  \bibinfo{journal}{Phys. Rev. B} \textbf{\bibinfo{volume}{79}},
  \bibinfo{pages}{132502} (\bibinfo{year}{2009}),
  \urlprefix\url{https://link.aps.org/doi/10.1103/PhysRevB.79.132502}.

\bibitem[{\citenamefont{Kusunose et~al.}(2011)\citenamefont{Kusunose, Fuseya,
  and Miyake}}]{kusunose2011puzzle}
\bibinfo{author}{\bibfnamefont{H.}~\bibnamefont{Kusunose}},
  \bibinfo{author}{\bibfnamefont{Y.}~\bibnamefont{Fuseya}}, \bibnamefont{and}
  \bibinfo{author}{\bibfnamefont{K.}~\bibnamefont{Miyake}},
  \bibinfo{journal}{Journal of the Physical Society of Japan}
  \textbf{\bibinfo{volume}{80}}, \bibinfo{pages}{054702}
  (\bibinfo{year}{2011}),
  \urlprefix\url{http://dx.doi.org/10.1143/JPSJ.80.054702}.

\bibitem[{\citenamefont{Fominov et~al.}(2015)\citenamefont{Fominov, Tanaka,
  Asano, and Eschrig}}]{FominovPRB2015}
\bibinfo{author}{\bibfnamefont{Y.~V.} \bibnamefont{Fominov}},
  \bibinfo{author}{\bibfnamefont{Y.}~\bibnamefont{Tanaka}},
  \bibinfo{author}{\bibfnamefont{Y.}~\bibnamefont{Asano}}, \bibnamefont{and}
  \bibinfo{author}{\bibfnamefont{M.}~\bibnamefont{Eschrig}},
  \bibinfo{journal}{Phys. Rev. B} \textbf{\bibinfo{volume}{91}},
  \bibinfo{pages}{144514} (\bibinfo{year}{2015}),
  \urlprefix\url{https://link.aps.org/doi/10.1103/PhysRevB.91.144514}.

\bibitem[{\citenamefont{Bergeret et~al.}(2001)\citenamefont{Bergeret, Volkov,
  and Efetov}}]{BergeretPRL2001}
\bibinfo{author}{\bibfnamefont{F.~S.} \bibnamefont{Bergeret}},
  \bibinfo{author}{\bibfnamefont{A.~F.} \bibnamefont{Volkov}},
  \bibnamefont{and} \bibinfo{author}{\bibfnamefont{K.~B.}
  \bibnamefont{Efetov}}, \bibinfo{journal}{Phys. Rev. Lett.}
  \textbf{\bibinfo{volume}{86}}, \bibinfo{pages}{4096} (\bibinfo{year}{2001}),
  \urlprefix\url{https://link.aps.org/doi/10.1103/PhysRevLett.86.4096}.

\bibitem[{\citenamefont{Bergeret et~al.}(2005)\citenamefont{Bergeret, Volkov,
  and Efetov}}]{bergeret2005odd}
\bibinfo{author}{\bibfnamefont{F.~S.} \bibnamefont{Bergeret}},
  \bibinfo{author}{\bibfnamefont{A.~F.} \bibnamefont{Volkov}},
  \bibnamefont{and} \bibinfo{author}{\bibfnamefont{K.~B.}
  \bibnamefont{Efetov}}, \bibinfo{journal}{Rev. Mod. Phys.}
  \textbf{\bibinfo{volume}{77}}, \bibinfo{pages}{1321} (\bibinfo{year}{2005}),
  \urlprefix\url{https://link.aps.org/doi/10.1103/RevModPhys.77.1321}.

\bibitem[{\citenamefont{Yokoyama et~al.}(2007)\citenamefont{Yokoyama, Tanaka,
  and Golubov}}]{yokoyama2007manifestation}
\bibinfo{author}{\bibfnamefont{T.}~\bibnamefont{Yokoyama}},
  \bibinfo{author}{\bibfnamefont{Y.}~\bibnamefont{Tanaka}}, \bibnamefont{and}
  \bibinfo{author}{\bibfnamefont{A.~A.} \bibnamefont{Golubov}},
  \bibinfo{journal}{Phys. Rev. B} \textbf{\bibinfo{volume}{75}},
  \bibinfo{pages}{134510} (\bibinfo{year}{2007}),
  \urlprefix\url{https://link.aps.org/doi/10.1103/PhysRevB.75.134510}.

\bibitem[{\citenamefont{Houzet}(2008)}]{houzet2008ferromagnetic}
\bibinfo{author}{\bibfnamefont{M.}~\bibnamefont{Houzet}},
  \bibinfo{journal}{Phys. Rev. Lett.} \textbf{\bibinfo{volume}{101}},
  \bibinfo{pages}{057009} (\bibinfo{year}{2008}),
  \urlprefix\url{https://link.aps.org/doi/10.1103/PhysRevLett.101.057009}.

\bibitem[{\citenamefont{Eschrig and L{\"o}fwander}(2008)}]{EschrigNat2008}
\bibinfo{author}{\bibfnamefont{M.}~\bibnamefont{Eschrig}} \bibnamefont{and}
  \bibinfo{author}{\bibfnamefont{T.}~\bibnamefont{L{\"o}fwander}},
  \bibinfo{journal}{Nature Physics} \textbf{\bibinfo{volume}{4}},
  \bibinfo{pages}{138} (\bibinfo{year}{2008}).

\bibitem[{\citenamefont{Linder et~al.}(2008)\citenamefont{Linder, Yokoyama, and
  Sudb\o{}}}]{LinderPRB2008}
\bibinfo{author}{\bibfnamefont{J.}~\bibnamefont{Linder}},
  \bibinfo{author}{\bibfnamefont{T.}~\bibnamefont{Yokoyama}}, \bibnamefont{and}
  \bibinfo{author}{\bibfnamefont{A.}~\bibnamefont{Sudb\o{}}},
  \bibinfo{journal}{Phys. Rev. B} \textbf{\bibinfo{volume}{77}},
  \bibinfo{pages}{174514} (\bibinfo{year}{2008}),
  \urlprefix\url{https://link.aps.org/doi/10.1103/PhysRevB.77.174514}.

\bibitem[{\citenamefont{Cr\'epin et~al.}(2015)\citenamefont{Cr\'epin, Burset,
  and Trauzettel}}]{crepin2015odd}
\bibinfo{author}{\bibfnamefont{F.~m.~c.} \bibnamefont{Cr\'epin}},
  \bibinfo{author}{\bibfnamefont{P.}~\bibnamefont{Burset}}, \bibnamefont{and}
  \bibinfo{author}{\bibfnamefont{B.}~\bibnamefont{Trauzettel}},
  \bibinfo{journal}{Phys. Rev. B} \textbf{\bibinfo{volume}{92}},
  \bibinfo{pages}{100507} (\bibinfo{year}{2015}),
  \urlprefix\url{https://link.aps.org/doi/10.1103/PhysRevB.92.100507}.

\bibitem[{\citenamefont{Yokoyama}(2012)}]{YokoyamaPRB2012}
\bibinfo{author}{\bibfnamefont{T.}~\bibnamefont{Yokoyama}},
  \bibinfo{journal}{Phys. Rev. B} \textbf{\bibinfo{volume}{86}},
  \bibinfo{pages}{075410} (\bibinfo{year}{2012}),
  \urlprefix\url{https://link.aps.org/doi/10.1103/PhysRevB.86.075410}.

\bibitem[{\citenamefont{Black-Schaffer and
  Balatsky}(2012)}]{Black-SchafferPRB2012}
\bibinfo{author}{\bibfnamefont{A.~M.} \bibnamefont{Black-Schaffer}}
  \bibnamefont{and} \bibinfo{author}{\bibfnamefont{A.~V.}
  \bibnamefont{Balatsky}}, \bibinfo{journal}{Phys. Rev. B}
  \textbf{\bibinfo{volume}{86}}, \bibinfo{pages}{144506}
  (\bibinfo{year}{2012}),
  \urlprefix\url{https://link.aps.org/doi/10.1103/PhysRevB.86.144506}.

\bibitem[{\citenamefont{Black-Schaffer and
  Balatsky}(2013{\natexlab{a}})}]{Black-SchafferPRB2013}
\bibinfo{author}{\bibfnamefont{A.~M.} \bibnamefont{Black-Schaffer}}
  \bibnamefont{and} \bibinfo{author}{\bibfnamefont{A.~V.}
  \bibnamefont{Balatsky}}, \bibinfo{journal}{Phys. Rev. B}
  \textbf{\bibinfo{volume}{87}}, \bibinfo{pages}{220506}
  (\bibinfo{year}{2013}{\natexlab{a}}),
  \urlprefix\url{https://link.aps.org/doi/10.1103/PhysRevB.87.220506}.

\bibitem[{\citenamefont{Triola et~al.}(2014)\citenamefont{Triola, Rossi, and
  Balatsky}}]{TriolaPRB2014}
\bibinfo{author}{\bibfnamefont{C.}~\bibnamefont{Triola}},
  \bibinfo{author}{\bibfnamefont{E.}~\bibnamefont{Rossi}}, \bibnamefont{and}
  \bibinfo{author}{\bibfnamefont{A.~V.} \bibnamefont{Balatsky}},
  \bibinfo{journal}{Phys. Rev. B} \textbf{\bibinfo{volume}{89}},
  \bibinfo{pages}{165309} (\bibinfo{year}{2014}),
  \urlprefix\url{https://link.aps.org/doi/10.1103/PhysRevB.89.165309}.

\bibitem[{\citenamefont{Cayao and Black-Schaffer}(2017)}]{cayao2017prb}
\bibinfo{author}{\bibfnamefont{J.}~\bibnamefont{Cayao}} \bibnamefont{and}
  \bibinfo{author}{\bibfnamefont{A.~M.} \bibnamefont{Black-Schaffer}},
  \bibinfo{journal}{Phys. Rev. B} \textbf{\bibinfo{volume}{96}},
  \bibinfo{pages}{155426} (\bibinfo{year}{2017}),
  \urlprefix\url{https://link.aps.org/doi/10.1103/PhysRevB.96.155426}.

\bibitem[{\citenamefont{Tanaka and Golubov}(2007)}]{tanaka2007theory}
\bibinfo{author}{\bibfnamefont{Y.}~\bibnamefont{Tanaka}} \bibnamefont{and}
  \bibinfo{author}{\bibfnamefont{A.~A.} \bibnamefont{Golubov}},
  \bibinfo{journal}{Phys. Rev. Lett.} \textbf{\bibinfo{volume}{98}},
  \bibinfo{pages}{037003} (\bibinfo{year}{2007}),
  \urlprefix\url{https://link.aps.org/doi/10.1103/PhysRevLett.98.037003}.

\bibitem[{\citenamefont{Tanaka et~al.}(2007)\citenamefont{Tanaka, Tanuma, and
  Golubov}}]{TanakaPRB2007}
\bibinfo{author}{\bibfnamefont{Y.}~\bibnamefont{Tanaka}},
  \bibinfo{author}{\bibfnamefont{Y.}~\bibnamefont{Tanuma}}, \bibnamefont{and}
  \bibinfo{author}{\bibfnamefont{A.~A.} \bibnamefont{Golubov}},
  \bibinfo{journal}{Phys. Rev. B} \textbf{\bibinfo{volume}{76}},
  \bibinfo{pages}{054522} (\bibinfo{year}{2007}),
  \urlprefix\url{https://link.aps.org/doi/10.1103/PhysRevB.76.054522}.

\bibitem[{\citenamefont{Linder et~al.}(2009)\citenamefont{Linder, Yokoyama,
  Sudb\o{}, and Eschrig}}]{LinderPRL2009}
\bibinfo{author}{\bibfnamefont{J.}~\bibnamefont{Linder}},
  \bibinfo{author}{\bibfnamefont{T.}~\bibnamefont{Yokoyama}},
  \bibinfo{author}{\bibfnamefont{A.}~\bibnamefont{Sudb\o{}}}, \bibnamefont{and}
  \bibinfo{author}{\bibfnamefont{M.}~\bibnamefont{Eschrig}},
  \bibinfo{journal}{Phys. Rev. Lett.} \textbf{\bibinfo{volume}{102}},
  \bibinfo{pages}{107008} (\bibinfo{year}{2009}),
  \urlprefix\url{https://link.aps.org/doi/10.1103/PhysRevLett.102.107008}.

\bibitem[{\citenamefont{Linder et~al.}(2010)\citenamefont{Linder, Sudb\o{},
  Yokoyama, Grein, and Eschrig}}]{LinderPRB2010_2}
\bibinfo{author}{\bibfnamefont{J.}~\bibnamefont{Linder}},
  \bibinfo{author}{\bibfnamefont{A.}~\bibnamefont{Sudb\o{}}},
  \bibinfo{author}{\bibfnamefont{T.}~\bibnamefont{Yokoyama}},
  \bibinfo{author}{\bibfnamefont{R.}~\bibnamefont{Grein}}, \bibnamefont{and}
  \bibinfo{author}{\bibfnamefont{M.}~\bibnamefont{Eschrig}},
  \bibinfo{journal}{Phys. Rev. B} \textbf{\bibinfo{volume}{81}},
  \bibinfo{pages}{214504} (\bibinfo{year}{2010}),
  \urlprefix\url{https://link.aps.org/doi/10.1103/PhysRevB.81.214504}.

\bibitem[{\citenamefont{Tanaka et~al.}(2012)\citenamefont{Tanaka, Sato, and
  Nagaosa}}]{TanakaJPSJ2012}
\bibinfo{author}{\bibfnamefont{Y.}~\bibnamefont{Tanaka}},
  \bibinfo{author}{\bibfnamefont{M.}~\bibnamefont{Sato}}, \bibnamefont{and}
  \bibinfo{author}{\bibfnamefont{N.}~\bibnamefont{Nagaosa}},
  \bibinfo{journal}{Journal of the Physical Society of Japan}
  \textbf{\bibinfo{volume}{81}}, \bibinfo{pages}{011013}
  (\bibinfo{year}{2012}), \eprint{http://dx.doi.org/10.1143/JPSJ.81.011013},
  \urlprefix\url{http://dx.doi.org/10.1143/JPSJ.81.011013}.

\bibitem[{\citenamefont{Parhizgar and
  Black-Schaffer}(2014)}]{parhizgar_2014_prb}
\bibinfo{author}{\bibfnamefont{F.}~\bibnamefont{Parhizgar}} \bibnamefont{and}
  \bibinfo{author}{\bibfnamefont{A.~M.} \bibnamefont{Black-Schaffer}},
  \bibinfo{journal}{Phys. Rev. B} \textbf{\bibinfo{volume}{90}},
  \bibinfo{pages}{184517} (\bibinfo{year}{2014}),
  \urlprefix\url{https://link.aps.org/doi/10.1103/PhysRevB.90.184517}.

\bibitem[{\citenamefont{Triola et~al.}(2016)\citenamefont{Triola, Badiane,
  Balatsky, and Rossi}}]{triola2016prl}
\bibinfo{author}{\bibfnamefont{C.}~\bibnamefont{Triola}},
  \bibinfo{author}{\bibfnamefont{D.~M.} \bibnamefont{Badiane}},
  \bibinfo{author}{\bibfnamefont{A.~V.} \bibnamefont{Balatsky}},
  \bibnamefont{and} \bibinfo{author}{\bibfnamefont{E.}~\bibnamefont{Rossi}},
  \bibinfo{journal}{Phys. Rev. Lett.} \textbf{\bibinfo{volume}{116}},
  \bibinfo{pages}{257001} (\bibinfo{year}{2016}),
  \urlprefix\url{https://link.aps.org/doi/10.1103/PhysRevLett.116.257001}.

\bibitem[{\citenamefont{Black-Schaffer and
  Balatsky}(2013{\natexlab{b}})}]{black2013odd}
\bibinfo{author}{\bibfnamefont{A.~M.} \bibnamefont{Black-Schaffer}}
  \bibnamefont{and} \bibinfo{author}{\bibfnamefont{A.~V.}
  \bibnamefont{Balatsky}}, \bibinfo{journal}{Phys. Rev. B}
  \textbf{\bibinfo{volume}{88}}, \bibinfo{pages}{104514}
  (\bibinfo{year}{2013}{\natexlab{b}}),
  \urlprefix\url{https://link.aps.org/doi/10.1103/PhysRevB.88.104514}.

\bibitem[{\citenamefont{Komendov\'a et~al.}(2015)\citenamefont{Komendov\'a,
  Balatsky, and Black-Schaffer}}]{komendova2015experimentally}
\bibinfo{author}{\bibfnamefont{L.}~\bibnamefont{Komendov\'a}},
  \bibinfo{author}{\bibfnamefont{A.~V.} \bibnamefont{Balatsky}},
  \bibnamefont{and} \bibinfo{author}{\bibfnamefont{A.~M.}
  \bibnamefont{Black-Schaffer}}, \bibinfo{journal}{Phys. Rev. B}
  \textbf{\bibinfo{volume}{92}}, \bibinfo{pages}{094517}
  (\bibinfo{year}{2015}),
  \urlprefix\url{https://link.aps.org/doi/10.1103/PhysRevB.92.094517}.

\bibitem[{\citenamefont{Komendov{\'a} and
  Black-Schaffer}(2017)}]{komendova2017odd}
\bibinfo{author}{\bibfnamefont{L.}~\bibnamefont{Komendov{\'a}}}
  \bibnamefont{and}
  \bibinfo{author}{\bibfnamefont{A.}~\bibnamefont{Black-Schaffer}},
  \bibinfo{journal}{arXiv preprint arXiv:1702.03181}  (\bibinfo{year}{2017}).

\bibitem[{\citenamefont{Triola and Black-Schaffer}(2018)}]{triola2018prb}
\bibinfo{author}{\bibfnamefont{C.}~\bibnamefont{Triola}} \bibnamefont{and}
  \bibinfo{author}{\bibfnamefont{A.~M.} \bibnamefont{Black-Schaffer}},
  \bibinfo{journal}{Phys. Rev. B} \textbf{\bibinfo{volume}{97}},
  \bibinfo{pages}{064505} (\bibinfo{year}{2018}),
  \urlprefix\url{https://link.aps.org/doi/10.1103/PhysRevB.97.064505}.

\bibitem[{\citenamefont{Triola and Balatsky}(2016)}]{triolaprb2016}
\bibinfo{author}{\bibfnamefont{C.}~\bibnamefont{Triola}} \bibnamefont{and}
  \bibinfo{author}{\bibfnamefont{A.~V.} \bibnamefont{Balatsky}},
  \bibinfo{journal}{Phys. Rev. B} \textbf{\bibinfo{volume}{94}},
  \bibinfo{pages}{094518} (\bibinfo{year}{2016}),
  \urlprefix\url{https://link.aps.org/doi/10.1103/PhysRevB.94.094518}.

\bibitem[{\citenamefont{Triola and Balatsky}(2017)}]{triola2017pair}
\bibinfo{author}{\bibfnamefont{C.}~\bibnamefont{Triola}} \bibnamefont{and}
  \bibinfo{author}{\bibfnamefont{A.~V.} \bibnamefont{Balatsky}},
  \bibinfo{journal}{Phys. Rev. B} \textbf{\bibinfo{volume}{95}},
  \bibinfo{pages}{224518} (\bibinfo{year}{2017}),
  \urlprefix\url{https://link.aps.org/doi/10.1103/PhysRevB.95.224518}.

\bibitem[{\citenamefont{Di~Bernardo
  et~al.}(2015{\natexlab{a}})\citenamefont{Di~Bernardo, Diesch, Gu, Linder,
  Divitini, Ducati, Scheer, Blamire, and Robinson}}]{di2015signature}
\bibinfo{author}{\bibfnamefont{A.}~\bibnamefont{Di~Bernardo}},
  \bibinfo{author}{\bibfnamefont{S.}~\bibnamefont{Diesch}},
  \bibinfo{author}{\bibfnamefont{Y.}~\bibnamefont{Gu}},
  \bibinfo{author}{\bibfnamefont{J.}~\bibnamefont{Linder}},
  \bibinfo{author}{\bibfnamefont{G.}~\bibnamefont{Divitini}},
  \bibinfo{author}{\bibfnamefont{C.}~\bibnamefont{Ducati}},
  \bibinfo{author}{\bibfnamefont{E.}~\bibnamefont{Scheer}},
  \bibinfo{author}{\bibfnamefont{M.~G.} \bibnamefont{Blamire}},
  \bibnamefont{and} \bibinfo{author}{\bibfnamefont{J.~W.}
  \bibnamefont{Robinson}}, \bibinfo{journal}{Nature communications}
  \textbf{\bibinfo{volume}{6}}, \bibinfo{pages}{8053}
  (\bibinfo{year}{2015}{\natexlab{a}}),
  \urlprefix\url{http://dx.doi.org/10.1038/ncomms9053}.

\bibitem[{\citenamefont{Di~Bernardo
  et~al.}(2015{\natexlab{b}})\citenamefont{Di~Bernardo, Salman, Wang, Amado,
  Egilmez, Flokstra, Suter, Lee, Zhao, Prokscha et~al.}}]{di2015intrinsic}
\bibinfo{author}{\bibfnamefont{A.}~\bibnamefont{Di~Bernardo}},
  \bibinfo{author}{\bibfnamefont{Z.}~\bibnamefont{Salman}},
  \bibinfo{author}{\bibfnamefont{X.~L.} \bibnamefont{Wang}},
  \bibinfo{author}{\bibfnamefont{M.}~\bibnamefont{Amado}},
  \bibinfo{author}{\bibfnamefont{M.}~\bibnamefont{Egilmez}},
  \bibinfo{author}{\bibfnamefont{M.~G.} \bibnamefont{Flokstra}},
  \bibinfo{author}{\bibfnamefont{A.}~\bibnamefont{Suter}},
  \bibinfo{author}{\bibfnamefont{S.~L.} \bibnamefont{Lee}},
  \bibinfo{author}{\bibfnamefont{J.~H.} \bibnamefont{Zhao}},
  \bibinfo{author}{\bibfnamefont{T.}~\bibnamefont{Prokscha}},
  \bibnamefont{et~al.}, \bibinfo{journal}{Phys. Rev. X}
  \textbf{\bibinfo{volume}{5}}, \bibinfo{pages}{041021}
  (\bibinfo{year}{2015}{\natexlab{b}}),
  \urlprefix\url{https://link.aps.org/doi/10.1103/PhysRevX.5.041021}.

\bibitem[{\citenamefont{Pivovarov and Nayak}(2001)}]{pivovarov2001odd}
\bibinfo{author}{\bibfnamefont{E.}~\bibnamefont{Pivovarov}} \bibnamefont{and}
  \bibinfo{author}{\bibfnamefont{C.}~\bibnamefont{Nayak}},
  \bibinfo{journal}{Phys. Rev. B} \textbf{\bibinfo{volume}{64}},
  \bibinfo{pages}{035107} (\bibinfo{year}{2001}),
  \urlprefix\url{https://link.aps.org/doi/10.1103/PhysRevB.64.035107}.

\bibitem[{\citenamefont{Kedem and Balatsky}(2015)}]{kedem2015odd}
\bibinfo{author}{\bibfnamefont{Y.}~\bibnamefont{Kedem}} \bibnamefont{and}
  \bibinfo{author}{\bibfnamefont{A.~V.} \bibnamefont{Balatsky}},
  \bibinfo{journal}{arXiv preprint arXiv:1501.07049}  (\bibinfo{year}{2015}).

\bibitem[{\citenamefont{Huang et~al.}(2015)\citenamefont{Huang, W\"olfle, and
  Balatsky}}]{huang2015odd}
\bibinfo{author}{\bibfnamefont{Z.}~\bibnamefont{Huang}},
  \bibinfo{author}{\bibfnamefont{P.}~\bibnamefont{W\"olfle}}, \bibnamefont{and}
  \bibinfo{author}{\bibfnamefont{A.~V.} \bibnamefont{Balatsky}},
  \bibinfo{journal}{Phys. Rev. B} \textbf{\bibinfo{volume}{92}},
  \bibinfo{pages}{121404} (\bibinfo{year}{2015}),
  \urlprefix\url{https://link.aps.org/doi/10.1103/PhysRevB.92.121404}.

\bibitem[{sm()}]{sm}
\bibinfo{howpublished}{Supplementary material.}

\bibitem[{\citenamefont{Yosida}(1958)}]{Yosida1958}
\bibinfo{author}{\bibfnamefont{K.}~\bibnamefont{Yosida}},
  \bibinfo{journal}{Phys. Rev.} \textbf{\bibinfo{volume}{110}},
  \bibinfo{pages}{769} (\bibinfo{year}{1958}),
  \urlprefix\url{https://link.aps.org/doi/10.1103/PhysRev.110.769}.

\end{thebibliography}

\end{document}